\documentclass[floatfix,twocolumn,pra,superscriptaddress,showpacs]{revtex4}

\usepackage{graphicx}
\usepackage{amsmath}
\usepackage{amssymb}
\usepackage{bm}
\usepackage{epstopdf}

\begin{document}

\title{Complex chaos in the conditional dynamics of qubits}

\author{T. Kiss}
\affiliation{Research Institute for Solid State Physics and
  Optics, H-1525 Budapest, P. O. Box 49, Hungary}
%\email{tkiss@szfki.hu}
\author{I. Jex}
\affiliation{Department of Physics, FJFI \v CVUT, B\v rehov\'a 7, 115 19 Praha 1 - Star\'e
M\v{e}sto, Czech Republic}
%\email{}
\author{G. Alber}
\affiliation{Institut f\"ur Angewandte
Physik, Technische Universit\"at Darmstadt, D-64289 Darmstadt,
Germany }
%\email{} 
\author{S. Vym\v etal}
\affiliation{Department of Physics, FJFI \v CVUT, B\v rehov\'a 7, 115 19 Praha 1 - Star\'e
M\v{e}sto, Czech Republic}

\date{\today}

\begin{abstract}
We analyse the consequences of iterative measurement-induced
non-linearity on the dynamical behaviour of qubits.  We present a
one-qubit scheme where the equation governing the time evolution is a
complex-valued non-linear map with one complex parameter.
In contrast to the usual notion of quantum chaos, exponential
sensitivity to the initial state occurs here. We calculate analytically 
the Lyapunov exponent based on the overlap of quantum states, 
and find that it is positive.   
We present a few illustrative examples of the emerging dynamics.
\end{abstract}


\maketitle

\flushbottom

Exponential sensitivity to inital conditions in nonlinear systems,
first described by Poincar\'e \cite{Poincare1892}, is today known as
chaos. Quantum systems which are classically chaotic generally do not
show exponential sensitivity even if their dynamics are complicated and
refered to as quantum chaos \cite{LesHouches91}. Unitarity of the
evolution generally prohibits exponential sensitivity, although rather
exotic exceptions were found \cite{exotic-chaos}.

Measuring a quantum system affects its dynamics.  As a result, instead
of the original unitary evolution an effective nonlinear dynamics may
emerge.  This phenomeon was investigated extensively in quantum
systems with continuous degrees of freedom which possess a
corresponding classical limit \cite{quantum-classical}.  In
particular, it was demonstrated that this sensitivity is strong in
dynamical regimes in which relevant actions are large in comparison
with Planck's constant and the corresponding classical dynamics are
chaotic.  The continuously monitored system evolves in such cases
according to a stochastic nonlinear Schr\"odinger equation, reflecting
the fact that measurements have non-deterministic output. Very recently,
Habib, Jacobs and Shizume could prove \cite{HJS} that a continuously measured classically chaotic system can be truly chaotic even far from any classical limit.
They numerically calculated the Lyapunov exponent and found that it is positive for the
expectation value of the position operator.

An alternative way of handling measurement results is to use them as
conditions and select the subensemble according to the prescribed output,
thus evoking a deterministic effective dynamics for the subensemble.
Conditional dynamics is of considerable current interest in the
context of quantum information science for qubit systems. In quantum
state purification protocols \cite{purification}, for example, such
non-linear effects are exploited to guarantee the unconditional
security of quantum cryptographic key distribution protocols
\cite{Briegel}.  Though basic aspects of measurement-induced
non-linear effects have already been investigated in these systems, it
is still largely unexplored whether these effects may lead to chaos
and, in case they do, what their characteristic dynamical features
are.

In this paper we investigate the dynamical features of iterated
deterministic quantum maps which describe the measurement-induced
conditional dynamics of one- and two qubit systems and which have been
proposed recently in the context of quantum information
\cite{Gisin98}. These qubit systems do not have a (trivial) classical
limit.  The idea that feeding results from weak measurements back
into the dynamics of an ensemble of quantum systems could possibly
lead to a novel type of quantum chaos with sensitivity to initial
conditions was mentioned already by Lloyd and Slotine
\cite{Lloyd2000}. The present scheme is related to this idea although
it is based not on weak but on standard strong selective measurements.
The selection, conditioned on measurement outcomes on part of the
system, can be thought of as feeding information back into the remaining 
subensemble.

As a main result it will be demonstrated that even in the simplest
possible case of one-qubit systems the resulting dynamics of pure
quantum states are governed by a special class of non-linear maps in
one complex-valued variable.  First studies of the iterative behavior
of such complex-valued non-linear maps were performed already a
century ago by Fatou \cite{Fatou06} thus paving the way for a new
research field within non-linear dynamics \cite{Milnor}.  Therefore, a
detailed understanding of the sensitivity of these measurement-induced
quantum maps with respect to initial states can be obtained by taking
advantage of the concepts and theorems from the so-far unrelated
theory of iterated maps with one complex variable. In order to
demonstrate the richness inherent in this 'complex chaos' a few
illustrative examples are presented.

Consider the nonlinear quantum transformation \cite{Gisin98}:
%%%
\begin{equation}
\rho ^\prime = {\cal S}\rho\, , \quad \rho_{ij} \xrightarrow{\cal S} N
\rho_{ij}^2\,
\label{maprho}
\end{equation}
%%%
with the renormalization factor $N=1/\sum\rho_{ii}^2$.  Thereby, the
matrix element squaring is defined with respect to a prescribed
orthonormal basis $\{|i\rangle\}$.  In the special case of qubits this
deterministic non-linear map can be realized by applying a
controlled-not gate on a pair of equally prepared qubits and then filtering by a
measurement performed on one of them \cite{Gisin98}.
By repeating the transformation ${\cal S}$ we expect that smaller
diagonal matrix elements will tend to zero and the largest one
survives, converging to unity. In other words, pure states can be stable
fixed points of the map, leading to purification of the
state. The maximally mixed state is also a fixed point of the map $\cal S$,
but it is easy to see that it is an unstable one. Perturbing the
initial state by increasing the weight of one of the states in the
mixture will lead to purification towards that particular state.

While the squaring operator itself behaves rather simply, the dynamics
become highly nontrivial if an additional local unitrary
transformation, a rotation in the qubit Hilbert space, is applied
%%%
\begin{equation}
\label{rotation1}
{\cal R} \rho = U \rho U^{\dagger} \, ,
\end{equation}
%%%
with 
%%%
\begin{equation}
U = \left(
\begin{array}{lll} 
&\cos x &  \sin x~e^{i\phi}\\
-&\sin x~e^{-i\phi} & \cos x
\end{array}
\right) \, ,
\label{rotation2}
\end{equation}
%%%
in the prescribed basis.  In this way one step of the dynamics reads
%%%
\begin{equation}
\rho^{\prime} = {\cal F} \rho = {\cal R S} \rho \, ,
\end{equation}
%%%
and repeating the transformation ${\cal F}$ leads to the discrete
conditional time-evolution we are interested in.
It should be kept in mind that each step of this particular
deterministic time evolution consists of a quantum mechanical filtering
process involving two qubits.  As a result of this filtering process
either the target qubit or both qubits have to be dismissed. 
The size of the original ensemble decreases exponentally in time.
%%%%%%%%%%%%%%%%%%%%%%%%%%%

Let us first consider an initial pure state of the qubit. With the notation 
%%%
\begin{equation}
|\psi\rangle = N (z |0\rangle + |1\rangle)\, ,
\end{equation}
%%%
where the state is normalized by $N=(1+|z|^{2})^{-1/2}$, the single
complex parameter $z$ describes the state of the qubit. The
transformation $\cal F$ maps this pure state onto a pure state and
transforms $z$ as
%%%
\begin{equation}
\label{mapz}
z \mapsto F_{p}(z)= \frac{z^{2}+p}{1-p^{*}z^{2}}\, ,
\end{equation}
%%%
where $p=\tan x e^{i \phi}$ and the star denotes complex conjugation.
The conditional dynamics of the qubit are thus governed by $F_{p}(z)$,
which is a non-linear $\mathbb{C}\to\mathbb{C}$ map with one complex
parameter $p$. A considerable difference compared to chaotic systems
in classical physics is that the underlying space is complex,
here. Even the simplest non-linear maps of the complex plane can show
intricate dynamical structure, such as the famous Mandelbrot set.  The
study of the mathematics related to maps in one complex variable has a
long history and an extensive literature, (for a review see
\cite{Milnor}). The map (\ref{mapz}) is a rational function of second
order polynomials, similar to the one first studied by Fatou
\cite{Fatou06} a century ago: $z \mapsto z^{2}/(z^{2}+2)$.

The traditional approach to a non-linear map in one complex variable
is to divide the complex plane of the initial values $z_{0}$ into
regular and irregular points forming the Fatou and Julia sets,
respectively. Regular starting points from the Fatou set will converge
to a stable cycle (also elements of the Fatou set) when repeating the
iteration. Taking into account both the initial condition $z_{0}$ and the
complex parameter $p$ a four dimensional parameter space is
defined. We will select special parameter values when studying the map
in order to illustrate the richness of its behaviour and demonstrate
sensitivity to the initial conditions.

The full dynamics induced by Eq. (\ref{mapz}) take place on the
Riemann sphere $\hat{\mathbb{C}}$ consisting of $\mathbb{C}$ together
with the point at infinity. The physical meaning of the points $0$ and
$\infty$ for $z$ are the two basis states of the qubit, $|1\rangle$
and $|0\rangle$, respectively. The map $F_{p}(z)$ is a rational
function of degree two. A general theorem on rational maps that are
quotients of two polynomials ensures that the Julia set is not empty. 
The non-vacuous Julia set is
the usual condition for complex valued maps to be considered chaotic. 
A more subtle question is whether the map is
hyperbolic, {\it i.e.} expanding on the Julia set. The latter property is closer 
to the sense how the term chaos is used for dynamical systems.   
For rational maps of degree two this can be decided by following the orbit of the
critical points: each orbit should converge to some attractive
periodic orbit. For definition and a review of the above properties we suggest to consult 
Milnor's book \cite{Milnor}.

The problem simplifies considerably if the parameter $p$ is set to
zero. The map is then the simple squaring $F_{0}(z)=z^{2}$. This is a
well-known example of a simple Julia set formed by the unit circle in
the complex plane. If the starting point is within the unit disk, the
iterations converge to zero, while initial values from outside
converge to infinity. Initial points with absolute value exactly one
will not converge, but the iterations follow an irregular path on the
unit circle. In other words, the qubit will tend to one of the basis
states, except for the equally weighted linear superposition. In the
latter case the relative phase will follow irregular dynamics. The
Julia set (a circle) is one-dimensional.
We use the following definition of the Lyapunov exponent $\lambda$
%%%
\begin{equation}
\label{ }
\lambda = \underset{n\to\infty}\lim \; \, \underset{\Delta(0)\to 0}\lim \;  \, \frac{1}{n} \ln \frac{\Delta(n)}{\Delta(0)} \, ,
\end{equation}
%%%
where we choose $\Delta$ to be the distance related to the overlap of the two corresponding quantum states:  $\Delta =1- | \langle \psi_{1} | \psi_{2} \rangle |^{2}$.
For unitary evolution this is a constant quantity, hence the above defined Lyapunov exponent of a closed quantum system is always zero.
It is generally not straightforward to apply this definition, since the result of the limit may depend on the path in the Hilbert space one takes when approaching the two initial states towards each other. Nevertheless, in our simple case we can restrict ourself to the unit circle. On this one dimensional manifold we may define the Lyapunov exponent with respect to the phase variable by choosing both initial states with $|z|=1$. Without loss of generality we can take $z_{0}=1\, ,\, z_{1}=e^{i \varphi}$ and arrive at
%%%
\begin{equation}
\label{Lyapunov-phi}
\lambda_{\varphi} = \underset{n\to\infty}\lim \; \, \underset{\varphi\to 0}\lim \;  \, \frac{1}{n} \ln \frac{\Delta_{\varphi}(n)}{\Delta_{\varphi}(0)} \, ,
\end{equation}
%%%
with $\Delta_{\varphi}(n) =  \frac{1}{2} (1-\cos 2^{n}\varphi)$. The order of limits is important and therefore, we can {\it first} let $\varphi$ go to zero, thus we can use the Taylor expansion of the cosine in $\varphi$ which leads to the result
\begin{equation}
\label{Lyapunov-result}
\lambda_{\varphi} =2 \ln 2 \, .
\end{equation}
%%%
The positive Lyapunov exponent indicates that exponentially fast separation takes place in the Hilbert space of the system.

Non-zero fixed values of $p$ are expected to lead to qualitatively
different dynamics. The corresponding Julia sets can possess highly
non-trivial structures.  By solving the $n$th order eigenvalue
equation $F^{\circ n}_{p}(z)=z$ (where $F^{\circ n}(z)$ denotes $n$
times repeated action of $F$), one can find various-order periodic
cycles $\{z_{1},z_{2}, \dots, z_{n}\}$, ($z_{i}\neq z_{j}$). The
stability of a cycle can then be determined by evaluating the
multiplier $\lambda=F_{p}^{\prime}(z_{1}) F_{p}^{\prime}(z_{2}) \dots
F_{p}^{\prime}(z_{n})$. If the absolute value of this multiplier is
smaller then unity, the fixed point is attractive. If it is greater
than unity it is repelling, while for $|\lambda|=1$ it is neutral.
Those neutral periodic cycles for which $\lambda$ is a root of unity,
and for which no iterate of the map is the identity are called
parabolic.  A rational map of degree two can have at most two cycles
which are attracting or neutral \cite{Milnor}.  While the first order
fixed points are given by a third order equation, in general the
period-$n$ fixed points would require the solution of a polynomial
equation of order $2^{n}+1$. The critical points $z_{c}$ of a map,
where $F^{\prime}(z_{c})=0$, play a special role in the theory of
non-linear maps.  For the map $F_{p}$ the two critical points are
$z_{c1}=0$ and $z_{c2}=\infty$, independent of $p$. The orbits of the
two critical points characterize a rational map to a large extent
\cite{Milnor93}. By checking the convergence of their orbit to an
attracting periodic cycle, one can decide whether the map is
hyperbolic. Moreover, all attractive and parabolic cycles can be found
in this way.

\begin{figure}
\includegraphics[width=7cm]{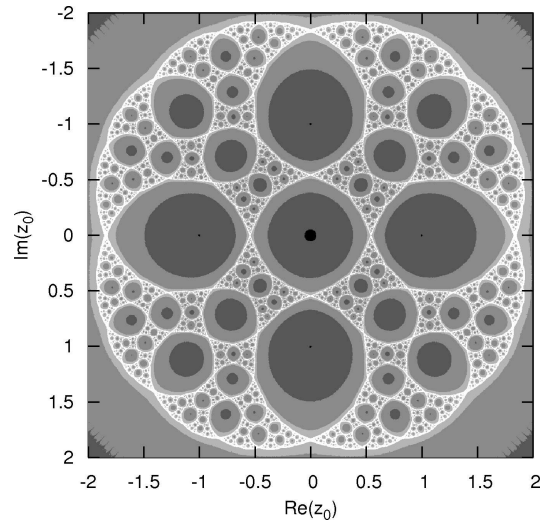}
\caption{ The Julia set for the non-linear map (\ref{mapz}). The
  parameter is set to $p=1$. Grayscale indicates how fast the map
  converges to the stable cycle $\{-1,\infty\}$ (dark -- fast, gray --
  slow convergence, white -- no convergence).}
\label{Fig-julia}
\end{figure}

Physically speaking, the parameter $p$ describes the rotation of the
qubit state $|\psi \rangle$. Setting the parameter to $p=1$
corresponds to a rotation of $\pi/4$ that transforms, for example, the
basis states into their equal superpositions.  This is a symmetric
situation with respect to the basis states. The orbit of one critical
point, $z_{c2}=\infty$, is part of the attractive cycle
$\{-1,\infty\}$. The other critical point $z_{c1}=0$ follows the orbit
$0 \mapsto 1 \mapsto \infty$ and thus lands on the same periodic
cycle.  Therefore the only stable cycle for this map is the fixed
point $\{-1,\infty\}$. This also proves that the map is hyperbolic
\cite{Milnor}.  Numerical calculation of the Julia set for quadratic
rational maps is difficult. There are no general algorithms to compute
it. Here we can simply apply the criterion of convergence to the
stable cycle when plotting the Julia set in Fig.(\ref{Fig-julia}) for
$p=1$. Dark points converge fast, gray points slower, white points do
not converge.  The complicated (fractal) structure of the Julia set
reflects the sensitivity of the dynamics to the initial state: a
change in the initial state may alter the dynamics from regular to
chaotic and this can occur on arbitrarily small scales.

\begin{figure}
\includegraphics[width=7cm]{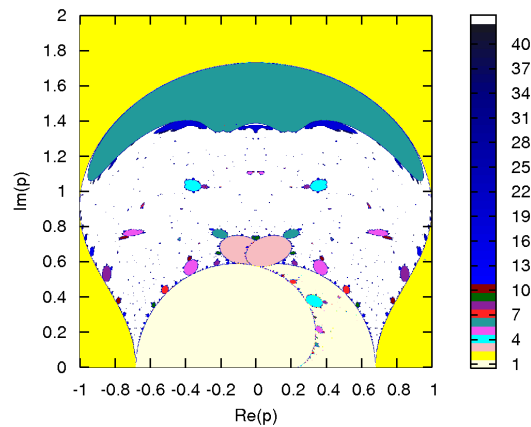}
\caption{ (Color online) The complex parameter space $p$ of
  the non-linear map (\ref{mapz}), with
  the initial state being the critical point $z_{0}=0$. Colors
  indicate the length of attractive cycles. White corresponds to no
  convergence. 
%  The outer region is formed by period-two fixed points
%  (orange). Period-one fixed points (yellow) concentrate around the
%  origin. The chaotic region (white) contains islands of various order
%  stable points.  
  }
\label{Fig-period}
\end{figure}

\begin{figure}
\includegraphics[width=6.6cm]{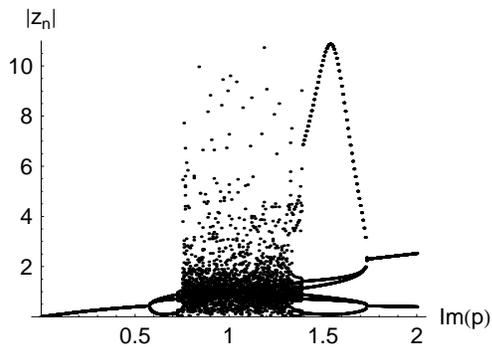}
\caption{Iterations of the non-linear map (\ref{mapz}) for purely
  imaginary $p$ with the initial state being the critical point
  $z_{0}=0$. After $10^{4}$ iterations the absolute values of $z$ for
  the next 50 steps are shown.  }
\label{Fig-itertable}
\end{figure}

The family of maps with varying values of $p \in \mathbb{C}$ may
possess fixed cycles of various length, but only at most two of them
can be attractive. The critical points converge to these attractive cycles, if their
orbit is convergent at all \cite{Milnor}. Fig. (\ref{Fig-period})
depicts the complex plane of $p$-values with colors showing the length
of the stable periodic cycle starting from the critical point
$z_{0}=0$. The lower half plane is not shown, since it is a mirror
image of the upper one.
The attractive periodic cycles are visualized for a fixed real value
of $p$ in Fig. (\ref{Fig-itertable}) by showing the absolute value of
$z$ after several iterations starting from $z_{0}=0$.

While the single qubit case well serves the purpose of rigorously
demonstrating the presence of complex chaos, two qubit systems with
conditional dynamics are of considerable practical interest for
quantum state purification. As the mathematical form of the procedure
of purification is similar in essence to the single-qubit case we
expect also similar dynamical properties for two-qubit systems. In
particular, parameter ranges and initial states should exist for which
purification protocols exhibit true chaos.
In order to address this question let us consider the following
iteration acting on two-qubit states \cite{Gisin98}
%%%
\begin{equation}
\label{twobitmap}
\rho^{\prime} = {\cal F} \rho = {\cal R}_1 {\cal R}_2 {\cal S} \rho \, .
\end{equation}
%%%
Thereby, ${\cal S}$ is the element squaring defined in
Eq. (\ref{maprho}) with the index $i$ running from 1 to 4 through the
elements of the product basis of the qubits $\{|j\rangle|k\rangle\}$,
$(j,k=0,1)$ and the rotation ${\cal R}_m$ acts on the $m$th qubit,
with parameters $x_m \, , \phi_m$ as defined in
Eqs. (\ref{rotation1},\ref{rotation2}).  Now, the parameters of the
two (local) complex rotations span $\mathbb{C}^{2}$, and the initial
state can be any valid two-qubit density operator. Obviously, this is
an even much larger parameter space to explore, which includes the
one-qubit pure states as a special case. Our numerical simulations indicate that 
meta-stable purification can occur here. An initial state with some deviation from a target pure state is being purified though several iterations, but then suddenly stability is lost
and chaos sets in \cite{inpreparation}. 

As an application, one could try to exploit the sensitivity of the system and use it
as a Schr\"odinger microscope \cite{Lloyd2000}. Tuning into a regime
where a few iterations amplify initial small differences of states one
could distinguish states exponentially fast. The cost paid is the
exponential size of the equally prepared systems needed.
To understand the general conditions which allow exponential sensitivity 
to initial states would be of use for any protocol applying measurement conditioned selection.

\acknowledgments{ Support by the Czech and Hungarian Ministries of
  Education (CZ-2/2005), by GA\v CR 202/04/2101, by MSMT LC 06001, by the European Union, by the Hungarian Scientific Research Fund (T043287 and T049234)  
  and by DAAD is acknowledged.
}


\begin{thebibliography}{20}

\bibitem{Poincare1892} H.~Poincar\'e, {\it Les M\'ethodes Nouvelles de
  la M\'echanique C\'eleste} (Gauthier-Villars, Paris, 1892).

\bibitem{LesHouches91} {\it Chaos and Quantum Physics}, Proceedings of
  the Les Houches Lecture Series, Session 52, eds. M.-J.~Giannoni,
  A.~Voros, and J.~Zinn-Justin (North-Holland, Amsterdam, 1991);
  P.~Cvitanovi\'c, R.~Artuso, R.~Mainieri, G.~Tanner and G.~Vattay,  {\it Chaos: Classical and Quantum}, 
{\tt ChaosBook.org} (Niels Bohr Institute, Copenhagen 2005).

\bibitem{exotic-chaos} B.V.~Chirikov {\it et al.}, Physica D {\bf 33}, 77 (1988); M.V.~Berry, 1992,
  in New Trends in Nuclear Collective Dynamics, eds:Y Abe, H Horiuchi
  and K Matsuyanagi (Springer proceedings in Physics 58), pp
  183-186.; R.~Bl\"umel, Phys. Rev. Lett. {\bf 73}, 428 (1994);
  R.~Schack, Phys. Rev. Lett. {\bf 75}, 581 (1995); R.~Bl\"umel,
  Phys. Rev. Lett. {\bf 75}, 582 (1995).

\bibitem{quantum-classical} R.~Schack {\it et al.},
  J. Phys. A: Math. Gen. {\bf 28}, 5401 (1995); T.~Bhattacharya {\it et al.}, Phys. Rev. Lett. {\bf 85}, 4852 (2000);
  A.J.~Scott and G.J.~Milburn, Phys. Rev. A {\bf 63}, 042101 (2001);
  G.G.~Carlo {\it et al.}, Phys. Rev. Lett. {\bf
    95}, 164101 (2005).

\bibitem{HJS} S.~Habib, K.~Jacobs, and K.~Shizume, Phys. Rev. Lett. {\bf 96}, 010403 (2006); S.~Habib {\it et al.}, e-print quant-ph/0505085.

\bibitem{purification} D.~Deutsch {\it et al}, Phys. Rev. Lett. {\bf 77},
  2818 (1996); C.~Macchiavello, Phys. Lett. A {\bf 246}, 385 (1998).

\bibitem{Briegel} H. Aschauer and H.J. Briegel,
  Phys. Rev. A {\bf 66}, 032302 (2002).

\bibitem{Gisin98} H.~Bechmann-Pasquinucci {\it et al.},
  Phys. Lett. A {\bf 242}, 198 (1998); D.R.~Terno, Phys. Rev. A {\bf
    59}, 3320 (1999); G.~Alber {\it et al.},
  J. Phys. A: Math. Gen. {\bf 34}, 8821 (2001).

\bibitem{Fatou06} P.~Fatou, C. R. Acad. Sci. Paris {\bf 143}, 546
  (1906).

\bibitem{Milnor} J.W.~Milnor {\it Dynamics in One Complex Variable},
  (Vieweg, 2000).

\bibitem{Lloyd2000} S.~Lloyd and J.-J.~Slotine Phys. Rev. A, {\bf 62},
  012307 (2000).

\bibitem{Milnor93} J.W.~Milnor, Exp. Math. {\bf 2}, 37 (1993).

\bibitem{inpreparation} T. Kiss, I. Jex and G. Alber (unpublished).

\end{thebibliography}
\end{document}